\documentclass[aps,showpacs,showkeys,preprint]{revtex4}
\usepackage{graphicx}
\usepackage{epsfig}
\usepackage{latexsym}
\usepackage{amsmath}

\begin{document}

\title{\boldmath Non-equilibrium phase and entanglement entropy in $2D$ holographic superconductors via Gauge-String duality}

\author{N.S. Mazhari}
  \email{najmemazhari86@gmail.com}
  \affiliation{Eurasian International Center for Theoretical Physics and Department of
General \& Theoretical Physics, Eurasian National University, Astana 010008, Kazakhstan}
\author{D. Momeni,\footnote{Corresponding author}}
  \email{momeni-d@enu.kz}
  \affiliation{Eurasian International Center for Theoretical Physics and Department of
General \& Theoretical Physics, Eurasian National University, Astana 010008, Kazakhstan}

\author{H. Gholizade}
  \email{hossein.gholizadehkalkhoran@tut.fi}
  \affiliation{Department of Physics, Tampere University of Technology  P.O.Box 692, FI-33101 Tampere, Finland}
\author{M. Raza}
  \email{mreza06@gmail.com}
  \affiliation{Department of Mathematics, COMSATS Institute of Information Technology, Sahiwal 57000, Pakistan}
\author{R. Myrzakulov}
  \email{rmyrzakulov@gmail.com}
  \affiliation{Eurasian International Center for Theoretical Physics and Department of
General \& Theoretical Physics, Eurasian National University, Astana 010008, Kazakhstan}

\date{\today}

\begin{abstract}
An alternative method of developing the theory of non-equilibrium two dimensional holographic superconductor is to start from the definition of a time dependent $AdS_3$ background. As originally proposed, many of these formulae were cast in exponential form, but the adoption of the numeric method of expression throughout the bulk serves to show more clearly the relationship between the various parameters.
 The time dependence behavior of the scalar condensation and Maxwell fields are fitted numerically. A usual value for Maxwell  field on AdS horizon  is $\exp(-bt)$, and the exponential $\log$ ratio is therefore $10^{-8} s^{-1}$. The coefficient $b$ of the time in the exponential term $\exp(-bt)$ can be interpreted as a tool to measure the degree of dynamical instability; its reciprocal $\frac{1}{b}$ is the time in which the disturbance is multiplied in the ratio. A discussion of some of the exponential formulae is given by the scalar field $\psi(z,t)$ near the AdS boundary.
It might be possible that a long interval would elapse  the system which tends to the equilibrium state when the normal mass and conformal  dimensions  emerged.  A somewhat curious calculation has been made, to illustrate the
 holographic entanglement entropy for this system. The foundation of all this calculation is, of course, a knowledge of multiple (connected and disconnected) extremal surfaces. There are several cases in which exact and approximate solutions are jointly used; a variable numerical quantity is represented by a graph, and the principles of approximation are then applied to determine related numerical quantities. In the case of the disconnected phase with a finite extremal are,  we find a discontinuity in the first derivative of the entanglement entropy as the conserved charge $J$ is increased.
\end{abstract}
\maketitle



\section{Introduction}
Statistical systems out of the equilibrium have been studied by non-equilibrium statistical mechanics. Recently inspired by the string theory, using a gauge-gravity duality, such
 non-equilibrium physics are revisited by holography\cite{CY,MKT,DNT,AJ,EH,Balasubramanian1,Balasubramanian2,HJW1,GPZ,BD,BBCCF,KKVT,HPG,HJW2,GS,CK,BLM,BGSSW,BDDN,
ACL,GGGZZ,BGS,CKY,Balasubramanian3,
BLMN,NNT,CKPT}. By holographic principle we mean a set up in which the physics of the gauge fields of the strongly coupled system can be extracted from the asymptotic behavior of the certain fields on the AdS boundary of gravitational bulk \cite{Maldacena}. Among all possibilities to consider bulk-boundary duality, lower dimensional models are very interesting because of their simplicities and analytical capabilities for investigations \cite{Liu:2011fy}-\cite{Nurmagambetov:2011yt}.\par
As an attempt to study non-equilibrium systems using AdS/CFT (gauge-string duality),in reference \cite{BDST}, the authors proposed a periodic driving model to confine themselves to discussing the holographic superconductor  and realistic-matter of it, indicating its relation to the varying chemical potential, and explaining the methods by which systems reach their condensation. In this connexion we may note that the equilibrium state  is reached at enough long times $t\to\infty$. Some critics propose to substitute for" dual chemical potential " $\mu$ the expression of $\rho(t)$ in a such manner that when $t\to\infty$, $\rho(t)\to\rho$. The method of attending the equilibrium varies in detail from state to state, but that most usual is for the dual parameters $\rho,\mu$ to propose time depending, often by a prescribed time dependent form, and for these situations, the equations of the motion to be solved on by the numerical methods. Besides the works already mentioned, \cite{Murata:2010dx},
 \cite{Li:2013fhw}, of which they may indicate the
 following:  considerations time dependent scalar and Maxwell fields; solving equations of the motion by numerical methods to find behavior of the fields near the AdS horizon to read the expectation values of the dual operators as functions of time. In this article we propose the same lower dimensional model, therefore to confine ourselves to discussing the character and subject-matter of the holographic superconductors, indicating its relation to $AdS_3/CFT_2$, and explaining the methods by which systems reach their equilibrium states, we don't assume any prescribed from for $\{\rho(t),\mu(t)\}$. A numerical algorithm written to propose a substitute for the system out of the equilibrium  is said to have fields $\psi(z,t),A_{\mu}(z,t)$. As long as they were time evolving together or disrupting the background, what difference did a small inhomogeneous field $A_x(z,t)$  makes  the system out of the equilibrium. There's a difference between this work and having the condensation with time dependent prescribed form for dual parameters.  The superconducting systems need an element to protect them from these out of the equilibrium fluctuations, not to mention the demons. In our article we stabilize the system using the inhomogeneous component of the field $A_x(z,t)$. It is commonly said that this is the difference between the equilibrium and the inequilibrium systems. The plan of this paper is as the following: In Sec. (\ref{Far-from}), we introduce the basic set up for such time dependence holographic superconductors. Equations of motion and dual AdS geometry is proposed. A gauge fixing is used to reduce the number  degree of freedoms of the system. In Sec. (\ref{Phase transition}), normal phase portrait is studied. In this section the system of the equations is simplified as much as we can. In Sec. (\ref{Numerical solutions}), we'll study numerical solutions for phases of system. Time dependence forms of the fields are investigated near the AdS boundary. In Sec. (\ref{Calculation of holographic}), we compute holographic entanglement entropy. We'll show that how we are able to calculate extremal surfaces in a time dependence background of bulk. We compute holographic entanglement entropy for connected and disconnected surfaces . In disconnected phase we calculate holographic entanglement for  two limits of the conserved charge $J$. We summarize in Sec. (\ref{Summary}).
\par
\section{Far-from-equilibrium model for $2D$ holographic superconductors }\label{Far-from}
Among the models of holographic superconductors it was undoubtedly \cite{Hartnoll:2008vx},
whose contributed most to the development of AdS/CFT based superconductors, and to its unfailing faith in their ultimate realization must be ascribed the completion of the first successful holographic model for strongly coupled system.
At the same time the bulk as a developed whole is regarded as an static and asymptotically AdS spacetime which is permeated with the AdS/CFT, and so we may say that the gravity bulk is a self-realization of the boundary world. The wellknown form of the  $AdS_3$ spacetime were already embedding  into the $4$D flat spacetime $\mathcal{R}^{2,2}$ with the following flat metric:
\begin{eqnarray}
g=-\big(dX_1^2+dX_2^2\big) +\big(dX_3^2+dX_4^2\big).\label{hyper}
\end{eqnarray}
We'll use  a set of the embedding coordinates
$\big(t,\rho,\theta\big)$ for all of the hyperboloids in this flat spacetime \cite {Nash:1956,Friedman:1961}:

\begin{eqnarray}
&&X_1=l\cosh\rho\sin t,\\&&
X_2=l\cosh\rho\cos t,\\&&
X_3=l\sinh\rho\sin\theta,\\&&
X_4=l\sinh\rho\cos\theta
\end{eqnarray}
Using this set of the coordinates,
and by speciefying an AdS radius
$l^2=-\frac{1}{\Lambda}$ we obtain the
universal covering  of the $AdS_3$ spacetime in the following static form:
\begin{eqnarray}
ds^2=l^2\big(-\cosh^2\rho dt^2+d\rho^2+\sinh^2\rho
d\theta^2\big)\label{unig}.
\end{eqnarray}
The domain $\rho=0$, represent the the AdS boundary of the AdS
metric. For our forthcoming desires to investigate the system out of the equilibrium, we need a non-static time dependent version of this metric.  It can be written in the infalling Eddington coordinates as
\begin{equation}
ds^2=\frac{l^2}{z^2}[-f(z)dt^2-2dtdz+dx^2]\label{metric1},
\end{equation}
 where the blackening factor is given by $f(z)=1-(\frac{z}{z_h})^2$ with $z=z_h$ the position of horizon and $z=0$ the AdS boundary.
\par

The formulation of $2D$ charged holographic superconductor with a global $U(1)$ symmetry was due to \cite{Hartnoll:2008vx}.
Let us start with the following bulk action in which gravity is coupled to an Abelian gauge field $A$ in the presence of a generally massive scalar field $\Psi$ with charge $q$ and mass $m$, i.e.,
\begin{eqnarray}
&&S=
\int_\mathcal{M} d^3x\sqrt{-g}[\frac{1}{2\kappa^2}\Big(R+\frac{2}{l^2}\Big)r+(-\frac{1}{4}F_{ab}F^{ab}-|D\Psi|^2-m^2|\Psi|^2)],
\end{eqnarray}
here
 $l$ is the AdS radius, and $D=\nabla-iA$ with $\nabla\equiv \partial_{\mu}$ , Maxwell tensor is $F_{\mu\nu}=\partial_{\mu}A_{\nu}-\partial_{\nu}A_{\mu}$. We'll work in the probe limit, which can be achieved by taking the  limit $\kappa^2=0$.
The equations of motion:
 \begin{eqnarray}
 &&D_\mu D^\mu\Psi-m^2\Psi=0,\\
&& \nabla_\mu F^{\mu\nu}=i[\Psi^* D^\nu\Psi-\Psi (D^\nu\Psi)^*].
 \end{eqnarray}
We define the CFT temperature as the Hawking temperature of the horizon by the following formula:
\begin{equation}
T=\frac{1}{2\pi z_h}\label{T}.
\end{equation}
The earliest formulation of the subject, due to \cite{Hartnoll:2008vx}, assumed that when we evaluate the Abelian  gauge field $A$ at the AdS boundary, it can be considered as a source with conserved current $J$. This later conserved current is  associated to a global $U(1)$ symmetry.
Furthermore,
the near AdS  boundary  $z\sim0$ data of the scalar field $\Psi$  provides  a  physical sources for the scalar operator $\mathcal{O}$. This relevant operator  has the conformal scaling dimension $\Delta_{\pm}=1\pm\sqrt{1+m^2l^2}$. In this article we study the case of $m^2l^2 =0$,$\Delta=\Delta_{+}=2$\\
With knowledge then of the conformal dimension  of scalar field of the conformal operators involved in boundary conformal action, we can at once calculate  the asymptotic solution of $A$ and $\Psi$  near the AdS boundary, by placing for each compound in the field equations its asymptotic metric, i.e $f(z)\sim0$. \\
From these numbers we can, by help of the  AdS/CFT dictionary, calculate the \emph{ vacuum expectation value (VEV) of the corresponding boundary quantum field theory operators}. In our case, these VEV quantities are  $\langle J\rangle$ and $\langle O\rangle$. Basically we can compute these quantities using the following variations of the effective and renormalized  action $\delta S_{ren}$ :
 \begin{eqnarray}
&& \langle J^\nu\rangle=\frac{\delta S_{ren}}{\delta a_\nu}=\lim_{z\rightarrow 0}\frac{\sqrt{-g}}{q^2}F^{z\nu},\\
&&\langle O\rangle=\frac{\delta S_{ren}}{\delta\phi}=\lim_{z\rightarrow 0}[\frac{z\sqrt{-g}}{lq^2}(D^z\Psi)^*-\frac{z\sqrt{-\gamma}}{l^2q^2}\Psi^*],
 \end{eqnarray}
 where $S_{ren}$ is the renormalized action :
 \begin{equation}
 S_{ren}=S-\frac{1}{lq^2}\int_\mathcal{B}\sqrt{-\gamma}|\Psi|^2,
 \end{equation}
and by dot we mean derivative with respect to the time coordinate  $t$. In our work for simplicity, we set
$l=1$, $q=1$, and $z_h=1$. We can calculate the expectation value of condensation of operators from its asymptotic fields for any substance lived in a given bulk, from a knowledge of the temperature  of boundary, by means of an application of the well-known thermodynamical process.
\par
Now we are fixing the fields as $\Psi=\Psi(x,z,t),{\bf A}=A_t(x,z,t)dt+A_{x}(x,z,t)dx$. By the fixing of this gauge field ${\bf A}$, the task of the time driven superconductor was considerably simplified.
The corresponding equations of motion are written in an explicit form:
\begin{eqnarray}
&&2\partial_t\partial_z\psi+\frac{2}{z^2}f\psi-\frac{f'}{z}\psi-f'\partial_z\psi-f\partial_z^2\psi-i\big(\partial_zA_t\psi+2A_t\partial_z\psi\big)
-\partial_x^2\psi+i\big(\partial_xA_x\psi+
2A_x\partial_x\psi\big)\nonumber\\
&&
+A_x^2\psi-\frac{2}{z^2}\psi=0
\end{eqnarray}
for the Klein-Gordon equation with $\Psi=z\psi$ and the following partial differential equations for gauge fields:
\begin{eqnarray}
&&\partial_z^2A_t-\partial_z\partial_xA_x=i(\psi^*\partial_z\psi-\psi\partial_z\psi^*),
\end{eqnarray}
\begin{eqnarray}
&&\partial_t\partial_zA_t+\partial_t\partial_xA_x
-\partial_x^2A_t-f\partial_z\partial_xA_x=-i(\psi^*\partial_t\psi-\psi\partial_t\psi^*)
-2A_t\psi^*\psi \\&&\nonumber+if(\psi^*\partial_z
\psi-\psi\partial_z\psi^*),
\end{eqnarray}
\begin{eqnarray}
&&\partial_z\partial_xA_t+f\partial_z^2A_x
+f'\partial_zA_x
-2\partial_t\partial_zA_x=i(\psi^*\partial_x\psi-\psi\partial_x\psi^*)+2A_x\psi^*\psi
\end{eqnarray}
By prime here we mean the differentiation  of the fields with respect to $z$. We suppose that the Maxwell gauge fields satisfy the following
 ansatz:
\begin{equation}
\partial_xA_t=\partial_xA_x=\partial_x\psi=0\label{ansatz}.
\end{equation}
Thus by equation (\ref{ansatz}), the equations of motion are expressed by :

\begin{eqnarray}
&&\label{eq1}2\partial_t\partial_z\psi+\frac{2}{z^2}f\psi-\frac{f'}{z}\psi-f'\partial_z\psi- f\partial_z^2\psi-i\partial_zA_t\psi-2iA_t\partial_z\psi+A_x^2\psi-\frac{2}{z^2}\psi=0,\\
&&\partial_z^2A_t=i(\psi^*\partial_z\psi-\psi\partial_z\psi^*),\label{eq2}\\
&&\partial_t\partial_zA_t=-i(\psi^*\partial_t\psi-\psi\partial_t\psi^*)
-2A_t\psi^*\psi+if(\psi^*\partial_z\psi-\psi\partial_z\psi^*),\label{eq3}\\
&&f\partial_z^2A_x+f'\partial_zA_x
-2\partial_t\partial_zA_x=2A_x\psi^*\psi\label{eq4}.
\end{eqnarray}
The preceding investigation for (\ref{eq1}-\ref{eq4}) is based upon the assumption that in passing from one section of the spacetime to another the homogeneity of the spacetime in the $x$ direction does not change.
\par
\section{Phase transition}\label{Phase transition}
 The main goal of this paper is to study such solutions of system (\ref{eq1}-\ref{eq4}) which they can describe superconductor phase. Before the condensation is started, the system is prepared in the normal phase in which the characteristic temperature of the system $T>T_c$, where $T_c$ denotes a specific critical temperature. We suppose that the system is prepared in the following normal (non superconductor) state:
\begin{equation}
A_x=\partial_t\psi=\partial_tA_t=0,
\end{equation}
If we apply this assumption to the field equations (\ref{eq1}-\ref{eq4}), we see that the
normal phases is static solution to the following equation of motion \begin{equation}
\partial_t\partial_z A_t=\partial^2_{zz}A_t=0
\end{equation}
Whose solution is given by the following:
 \begin{equation}
\psi=0, A_t=\mu(1-z).
 \end{equation}
here by
$\mu$ we denote the dual chemical potential of the boundary system. To investigate system in the critical phase $T<T_c$  and when the system enters a phase transition region from normal to superconductor phase, we need to
 gauge fixing  as $\psi=\psi^*$. With this gauge fixing we must solve the following system of partial differential equations (PDEs):
\begin{eqnarray}
&&\label{pde1}2\partial_t\partial_z\psi+\frac{2}{z^2}f\psi-\frac{f'}{z}\psi-f'\partial_z\psi-f\partial_z^2\psi-i\partial_zA_t\psi-2iA_t\partial_z\psi+A_x^2\psi-\frac{2}{z^2}\psi=0,\\
&&\partial_z^2A_t=0,\label{pde2}\\
&&\partial_t\partial_zA_t+2A_t\psi^2=0\label{pde3}\\&&
f\partial_z^2A_x+f'\partial_zA_x
-2\partial_t\partial_zA_x-2A_x\psi^2=0\label{pde4}.
\end{eqnarray}
here $\psi=\psi(z,t), A_t=A_t(z,t),A_x=A_x(z,t)$.
\par
\section{Numerical solutions}\label{Numerical solutions}
\begin{center}
\begin{figure}[htb]
\epsfxsize=7cm \epsfbox{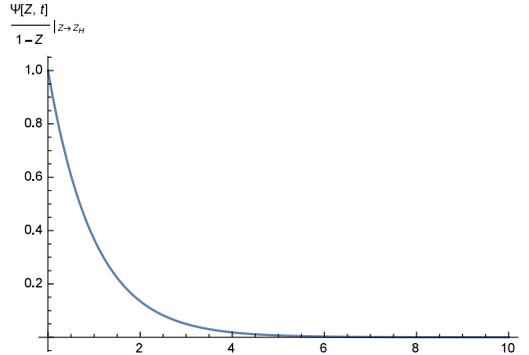} \caption{ $\frac{\psi(z,t)}{1-z}$ near AdS boundary point.}\label{f1}
\end{figure}
\end{center}

The ultimate value of numerical solutions must depend on the boundary conditions on which they are based. The value of the gauge field , for instance, is the result of many factors, some inherent, some due to boundary conditions, and until these have been sifted out, numerical methods of harmonics or of correlation can have no more than an empirical value. The appropriate boundary condition near the
 AdS horizon $z=0$ can be written as the following:
 \begin{equation}
 A_t(0,t)=0, \psi(0,t)=\langle O_{+}(t)\rangle z^2 .
 \end{equation}
In the vicinity of the horizon $z\simeq1$, the appropriate set of
 BCs are given by the following:
 \begin{equation}
\partial_z A_t=-\rho(t),\psi(z=1,t)=\Sigma_{n=1}^{\infty} \psi^{(n)}(z=1,t)(1-z)^n.
 \end{equation}
Such a hypothetical simplicity is the necessary step for solving the system of PDEs given in (\ref{pde1}-\ref{pde4}).  Then by solving these equations, regarding the four elements $\psi=\psi(z,t), A_t=A_t(z,t),A_x=A_x(z,t)$ as unknown quantities, the values of the l$\{\langle O_{+}(t)\rangle, \psi^{(n)}(z=1,t)\}$  may be computed.  we find that, eliminating  (\ref{pde2}), the resultant is a homogeneous function of  $z$ of degree $1$;
\begin{eqnarray}
&&A_t(z,t)=\mu(t)-\rho(t)z\label{sol-pde2}
\end{eqnarray}
differentiating $\partial_z$ of (\ref{pde3}) and solving for $\psi(z,t)$ we obtain the scalar field profile;
\begin{eqnarray}
&&\psi(z,t)=\sqrt{\frac{g(t)}{\mu(t)-\rho(t)z}}\label{sol-pde3}
\end{eqnarray}

 if values of $g(t),\mu(t),\rho(t)$ , given by any solution, be substituted in each of the two equations, they will possess a common factor which gives a value of $A_x(z,t)$ , yields a system of values which satisfies both equations:

\begin{eqnarray}
\Big[f\partial_z^2+f'\partial_z
-2\partial_t\partial_z-2\Big(\frac{g(t)}{\mu(t)-\rho(t)z}\Big)^2\Big]A_x(z,t)=0\label{pde4-1}
\end{eqnarray}
and we can solve (\ref{pde1},\ref{pde4-1}).
When $t\to \infty$ the system tends to the equilibrium superconductor phase.
Accordingly the graph of  (\ref{f1}) for $\frac{\psi(z,t)}{1-z}$, time evolution of the factor $g(t)$ in (\ref{sol-pde3}) is exponentially decreasing function of time.
We assumed that the values of chemical potential $\mu$ and charge density $\rho$, at each point, was slightly modified, from the equilibrium forms  belonging to a uniform $\mu\sim\rho$, by the time dependent factor $g(t)$ in the  (\ref{sol-pde3}).  The \emph{rate of diminution of amplitude} $\langle O_{+}(t)\rangle$ expressed by the coefficient $g(t)$ in the (\ref{sol-pde3}) is decreasing by time. The coefficient $b$ of the time in the exponential term $\frac{\psi(z,t)}{1-z}\sim g(t)$ can be used to measure the degree of dynamical (in)stability; its reciprocal $\frac{1}{b}$ is the time in which the disturbance is multiplied in the ratio i. e. $\tau\sim 10^8$.

\begin{figure}[htb]
\epsfxsize=7cm \epsfbox{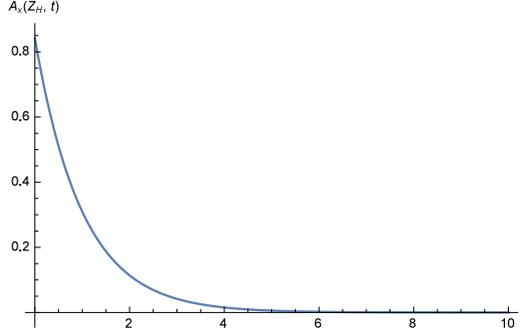} \caption{ $A_x(t)$ near AdS boundary point.}\label{axzh}
\end{figure}

\emph{As numerical results indicate, after long time the $A_x$ tends to zero (cf. figure (\ref{axzh})). Therefore, we can ignore the effects of $A_x$ in the equations and solve the equations of motion.} The analytical expression for the $A_x$ (\ref{pde4-1}) in the latter case involves exponential terms, one of which decreases rapidly, being equally multiplied in equal times:
\begin{eqnarray}
&&A_x \left( z,t \right) ={\it F_2} \left( t \right)- \frac{1}{2}\,{\it F_1}
 \left( 2\,{\tanh}^{-1} \left( z \right) +t \right)
\end{eqnarray}
When $t\to\infty$ and near the AdS boundary point, we should satisfy an auxiliary condition ${\it F_2} \left( t \right)- \frac{1}{2}\,{\it F_1}
 \left( t \right)|_{t\to\infty}=0$.

Within this approximation, the result for $g(t)$ is:
\begin{eqnarray*}
  g(t) &\sim &  \exp \left(-t\right)
\end{eqnarray*}


Near $z_h$ we can analyse the numerical result without setting the $A_x=0$. The figure (\ref{f1}) shows that the form of $g(t)$ near the AdS boundary $z=0$.
The numerical non-linear fitting results  for $g(t)$ confirm analytical results:
\begin{eqnarray}
  g(t) &=& a \exp(-b t), \\
   a &=& 1.10517 \pm 1.95684\times10^{-8}, \\
   b &=& 1.0 \pm 2.37993\times10^{-8},
\end{eqnarray}
The coefficient $a$ of this function $g(t)$ is equivalent to $\langle O_{+}(t)\rangle,\ \ t\gg\tau$. The coefficient of $\langle O_{+}(t)\rangle|$ is a variable quantity depending upon the temperature $T$ of thesystem given in (\ref{T}) , but is usually taken to be $\sqrt{1-\frac{T}{T_c}}$ is the fundamental equation between the condensation and temperature, however the lower order terms may be applied which all vanishes at critical point $T=T_c$.


\begin{figure}[htb]
\epsfxsize=7cm \epsfbox{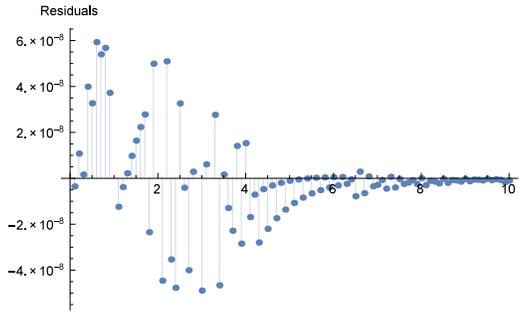} \caption{ Fitting residuals near $z\sim0$ for $g(t)$.}\label{rezh}
\end{figure}

Perhaps the best data for a comparison are those afforded by the computing  fitting residuals for $g(t)$  at different times.
Figure (\ref{rezh}) shows the fitting residuals for $g(t)$ near AdS boundary.

\begin{figure}[htb]
\epsfxsize=7cm \epsfbox{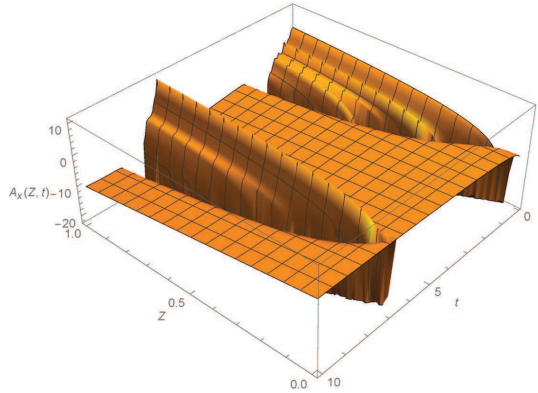} \caption{ $3$D plot of $A_x(z,t)$ as a solution for (\ref{pde4-1}). Furthermore, we observe a discontinuity in $A_x(0,t)$ near critical point $T_c=0.01$ at time scale $t_{f}\sim 7,t_{i}\sim 0.5$. For time interval $t\in[t_i,t_f]$, it shows that $A_x(z,t)$ is a monotonic-constant function.}\label{f4}
\end{figure}
Then by solving  equation (\ref{pde4-1}), regarding the boundary conditions  , the value of the  $A_x(z,t)$ may be computed in Fig.(\ref{f4}). From $t=0$ to $t\sim5$ we see a symmetric pattern which will be copied in the next interval times. On Fig. (\ref{f4}) , in the vicinity $z=0$, are the remains of an decreasing exponential factor $g(t)$, said to have been raised by $\psi(z,t)$ in (\ref{pde4-1}).  For time interval $t\in[t_i,t_f]$,it shows that $A_x(z,t)$ is a monotonic-constant function.

\begin{figure}[htb]
\epsfxsize=7cm \epsfbox{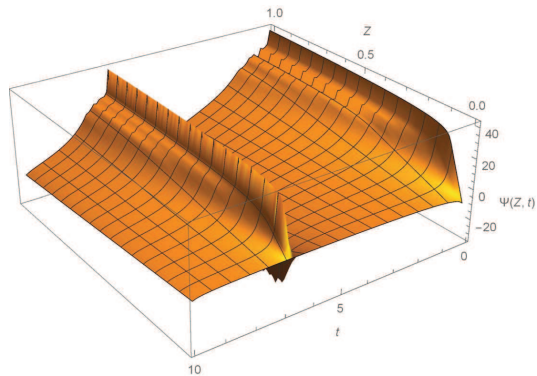} \caption{ 3D $\psi(z,t)$ as a solution for (\ref{pde1}). We observe a discontinuity in $\psi(0,t)$ near critical point $T_c=0.01$ at time scale $t\sim 7$.  For time interval $0<t< 7, 7<t<10$,it shows that $\psi(z,t)$ is a monotonic-decreasing function.}\label{f5}
\end{figure}
We found a temporary numeric solution to (\ref{pde1}), and we plotted the $\psi(z,t)$ around the $(t,z)$ in Fig.(\ref{f5}). An exponential decreasing form is observed. Furthermore, we observe a discontinuity in $\psi(0,t)$ near critical point $T_c=0.01$ at time scale $t\sim 7$. For time intervals $0<t< 7, 7<t<10$,it shows that $\psi(z,t)$ is a monotonic-decreasing function.

\par
\section{Calculation of holographic entanglement entropy }\label{Calculation of holographic}
The holographic entanglement entropy (HEE) of a quantum system in boundary is defined as the entropy of a region of space $\tilde{A}$ and its complement  on the minimal surfaces in $AdS_{d+1}$ using gauge-gravity duality \cite{hee1}-\cite{Nishioka:2009un}:

\begin{eqnarray}
S_{\tilde{A}}\equiv S_{HEE}=\frac{Area(\gamma _{\tilde{A}})}{4G_{d+1}}.\label{HEE}
\end{eqnarray}%
For time-independent backgrounds we need to compute the minimal area of a region in bulk with the same boundary $\partial A$ with the quantum system in boundary. This idea provided a very useful framework to study phase transitions in strongly coupled systems, specially in holographic superconductors
\cite{Peng:2014ira}-\cite{Romero-Bermudez:2015bma}.

Formalism for time dependence backgrounds  presented in \cite{Hubeny:2007xt}, it is recently used in time-dependent Janus background \cite{Ugajin:2014nca}. The technique is to replace "minimal" with "extremal" surfaces:
\begin{eqnarray}
S^{time-dependent}_{\tilde{A}}\equiv S_{HEE}=ext\Big[\frac{Area(\gamma _{\tilde{A}})}{4G_{d+1}}\Big].\label{HEE}
\end{eqnarray}
In case of multiple extremal surfaces, we should select the extremal surface with the minimum area included in them.\par
In applying gauge-string techniques, we use another time coordinate $y=t+2\tanh^{-1} z$ in (\ref{metric1}), because the metric on AdS boundary becomes flat which is essential for CFT. The new appropriate form of metric is given by the following:
\begin{eqnarray}
&&
ds^2=\frac{l^2}{\tanh^2\Big(\frac{t-y}{2}\Big)}\Big[dx^2-\cosh^{-2}\Big(\frac{t-y}{2}\Big)dtdy\Big]\label{metric2}.
\end{eqnarray}
 In this coordinate $y$, the black hole horizon $z=1$ corresponds to $y = \infty$. Furthermore, the conformal (AdS) boundary $z=0$ is located at the sheet $y-t=0$. So, the metric on conformal boundary is manifested as flat one $ds^2\sim dx^2-dt^2$. \par
To compute the extremal surfaces, we should select an appropriate setup with the subsystem $A = \{(\pm y_{\infty}, t_{\infty}, x)| -x_{\infty} \leq x \leq x_{\infty}\}$ in the  bulk geometry (\ref{metric2}). Basically we have   two types of the extremal surface which are different from topologies. These types of extremal surfaces are defined by  connected phase and disconnected phase. The functional which should be extremized is:
\begin{eqnarray}
&&\label{func}A[t(y),x(y)]=l\int_{-y_{\infty}}^{y_{\infty}}\frac{dy}{\tanh\Big(\frac{t-y}{2}\Big)}\sqrt{-\dot{t}\cosh^{-2}\Big(\frac{t-y}{2}\Big)+\dot{x}^2}.
\end{eqnarray}
here dot denotes
 the time derivative $\frac{d}{dy}$. The problem is reduced to solve the Euler-Lagrange (EL) equation(s).
\subsection{Extremal areas in connected phase}
For connected surfaces, the surfaces are defined at $x=\pm x_{\infty}$, so the functional reduces to the following form:
\begin{eqnarray}
&&\frac{A[t(y)]}{l}=\int_{-y_{\infty}}^{y_{\infty}}\frac{dy}{\sinh\Big(\frac{t-y}{2}\Big)}\sqrt{-\dot{t}}\label{func2}.
\end{eqnarray}
There is no conserved quantity  associated to the coordinate $t$. We should find the solution of the following EL equation in which we introduced a new function $u(y)=\frac{t(y)-y}{2}$:
\begin{eqnarray}
&&\frac{d^2u(y)}{dy^2}=\coth(u(y))(1+\frac{du(y)}{dy})(1+2\frac{du(y)}{dy})\label{ode1}
\end{eqnarray}
This differential equation has the following first integral solution:
\begin{eqnarray}\label{huf1}
&&\label{sol-ode1}(1+\dot{u})\sqrt{-(1+2\dot{u})}=c\sinh u.
\end{eqnarray}
This expression gives the unique exact solution of
the equations of motion (\ref{ode1}).

We can compute explicitly the total area of the connected extremal surface  by the following functional:
\begin{eqnarray}
&&\frac{A[t(y)]}{l}=\int_{-y_{\infty}}^{y_{\infty}}\frac{dy}{\sinh\Big(\frac{u(y)}{2}\Big)}\sqrt{-(1+2\dot{u})}\label{func3}.
\end{eqnarray}
Using equation(\ref{huf1}) we can rewrite the equation(\ref{func3}) as:

\begin{eqnarray*}
\frac{A[t(y)]}{l}=\int_{u(-y_{\infty})}^{u(y_{\infty})} \frac{\cosh(\frac{u}{2})}{(\xi(u)+1) \xi(u)}du,\ \
\xi(u)= \frac{1}{6} (\delta +\frac{1}{\delta}-5)
\end{eqnarray*}
here $\delta=\sqrt[3]{-27 c^2 \cosh (2 u)+3 \sqrt{6} \sqrt{c^2 \sinh ^2(u) (27 c^2 \cosh (2 u)-27 c^2-2)}+27 c^2+1}$.

We should notice that we take the real solution of equation (\ref{huf1}).
Expanding $\frac{A[t(y)]}{l}$ in powers of $c$, we have:

\begin{eqnarray*}
  \frac{A}{l} = \int_{u(-y_{\infty})}^{u(y_{\infty})}
  du\Big( -196608 c^{10}(\sinh ^{11}(u) csch(\frac{u}{2}))-9728 c^8(\sinh ^9(u) csch(\frac{u}{2}))\\
  -512 c^6 (\sinh ^7(u) csch(\frac{u}{2})) -64 c^4(\sinh ^4(u) \cosh (\frac{u}{2})) -4 \cosh (\frac{u}{2})\Big)
\end{eqnarray*}

\begin{figure}[htb]
\epsfxsize=7cm \epsfbox{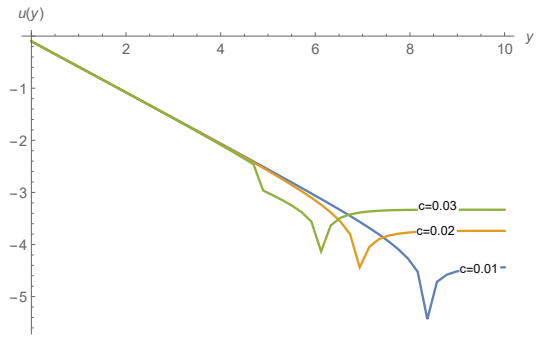} \caption{ Plot of the solution $u(y)$ as a function of $y$ for different values of $c=0.01, 0.02, 0.03$.}
\end{figure}

\begin{figure}[htb]
\epsfxsize=7cm \epsfbox{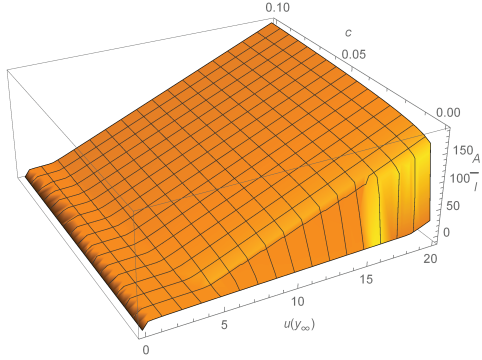} \caption{ Plot of the surface
$\frac{A}{l}$ versus $u(y_{\infty})$ and $c$. When $y$ is increasing, $u(y)$ in decreasing up to a local minima, after that it is increasing by a roughly slope. The minima is located at the turning point.  \emph{If we're increasing $c$, the local minima is shifted to the right. It can be interpreted as a shift in the horizon.} }
\end{figure}
In figure (6) we  numerically constructed $u(y)$ as a function of $y$ for different values of $c=0.01, 0.02, 0.03$. When $y$ is increasing, $u(y)$ in decreasing up to a local minima, after that it is increasing by a roughly slope. The minima is located at the turning point.  If we're increasing $c$, the local minima is shifted to the right. It can be interpreted as a shift in the horizon. \par
In figure (7) we plot regularized HEE per length $l$ as a function of the $\{u(y_{\infty}),c\}$.  It shows that $\frac{A}{l}$ is a
monotonic-increasing function.  It always increasing or remaining
constant, and never decreasing. It produces a regular phase of
matter for $T>T_c$ or equivalently for $u(y_{\infty})\succeq15.5$. Regular attendance at these non superconducting
phase has proved numerically. Boundary conditions and regular tiny
parameter $c$ will help to keep normal phase for longer.
Normal phase increasing the entropy $\frac{A}{l}$,  increases the
hardenability of superconductivity.

\subsection{ Extremal areas in disconnected phase}
In disconnected phase, we should compute the extremal area of the disconnected surfaces as a function of boundary coordinates $(t_{\infty},x_{\infty})$. We solve the EL equations for $t(y)$ and $x(y)$ for general functional (\ref{func}). The associated Noether charge for $x$ is given by $J\equiv \frac{\partial }{\partial\dot{x}}\Big(\frac{A[t(y),x(y)]}{l}\Big)$. This conserved charge $J$ plays a crucial role in our forthconing results.

If we solve it for $\dot{x}$ and substitute the result in (\ref{func}) we obtain:
\begin{eqnarray}
&&\label{func4}A[u(y)]= l\int_{-y_{\infty}}^{y_{\infty}}\frac{dy \sqrt{-(1+2\dot{u})}}{\sinh u(y)\sqrt{1-4J^2\tanh^2u(y)}}.
\end{eqnarray}
The returning point location $u_{*}\equiv(y_{*}, t_{*})$ is the one which $\dot{x}\to \infty$.  So, $J=\frac{1}{2\tanh u_{*}}$. The EL equation for $u(y)$ in (\ref{func4})   is given by:
\begin{eqnarray}
&&\ddot{u}=\frac{(1+2\dot{u})(\dot{u}^2-2\dot{u}-1)((4J^2-1)\cosh^4u-4J^2)}{(1+\dot{u})\sinh(2u)\Big(2J^2\sinh^2u-\frac{\cosh^2u}{2}\Big)}\label{ode2}.
\end{eqnarray}
This differential equation  has a unique first integral solution:
\begin{eqnarray}
&&\label{sol-ode2}\ln\sqrt{-(1+2\dot{u})(\dot{u}^2-2\dot{u}-1)}=\frac{
\Sigma_{i=0}^5 h_{i}(u,J)J^i}{12\tanh^3 u},
\end{eqnarray}
whose general solution is given by the following list of functions:

\begin{table}
\caption{Coefficents $h_i(u,J)$ in (\ref{sol-ode2})
.}
\begin{center}
\begin{tabular}{cc}
\hline\hline
~i& $h_i(u,J)$ \\
0&$- \frac{1}{\cosh^{6}( \frac{1}{2}u )}$ \\
1& $24\tanh^{3}( \frac{1}{2}u )\ln\Big(\frac{\tanh^{2}( \frac{1}{2}u )+4J\tanh( \frac{1}{2}u )+1}{-\tanh^{2}( \frac{1}{2}u )+4J\tanh( \frac{1}{2}u )-1}\Big)$\\
2&$\frac{4(1+\tanh^{4}( \frac{1}{2}u) -10\tanh^{2}( \frac{1}{2}u ))}{\cosh^{2}( \frac{1}{2}u )}$ \\
3  &$4h_1(u,J)$\\
4& $\frac{192\tanh^{2}( \frac{1}{2}u )}{\cosh^{2}( \frac{1}{2}u )}$\\
5 & $-16h_1(u,J)$\\
\hline\hline
\end{tabular}
\end{center}
\end{table}

Equation (\ref{sol-ode2}) has three solutions for $\dot{u}$. We can summarize the solutions ($\dot{u}-\frac{1}{2}$) as:

\begin{align*}
\dot{u}-\frac{1}{2}=\frac{\zeta}{6 \sqrt[3]{2}}
+\frac{11}{2^{2/3} \zeta}
-\frac{\left(1-i \sqrt{3}\right) \zeta}{12 \sqrt[3]{2}}-\frac{11 \left(1+i \sqrt{3}\right)}{2\ 2^{2/3}\zeta}
-\frac{\left(1+i \sqrt{3}\right) \zeta}{12 \sqrt[3]{2}}-\frac{11 \left(1-i \sqrt{3}\right)}{2\ 2^{2/3} \zeta}
\end{align*}

here
\begin{eqnarray}
&&\theta=\frac{\sum _{i=0}^5 h_i J^i}{12 \tan ^3(u)},\ \ \zeta=\sqrt[3]{\sqrt{\left(378-108 e^{2 \theta}\right)^2-143748}-108 e^{2\theta}+378}
\end{eqnarray}
We select the first  (real) solution for $\dot{u}$. We can investigate some limits of extremal surface areas in disconnected phase:
  \subsubsection{ Near $J=\frac{1}{2}$}
The case $J=\frac{1}{2}$ is easy to check that the only possible solution for (\ref{ode2}) is given by $\dot{u}=u_0$. If we compute (\ref{func4}) for this solution, we obtain:
\begin{eqnarray}\label{51}
\frac{A}{l}=2\sqrt{-(1+2u_0)}y_{\infty}
\end{eqnarray}
If system evolves in the vicinity of $J=\frac{1}{2}$, a numerical computation shows that  $\frac{A}{l}$ is a
monotonic-decreasing function.  Near $J_c\approx 0.58^{+}_{-}0.02$ we can simplify $\dot{u}=T_1+T_2(J-0.5)$ where $T_i$ are functions of $u$. The expressions $T_i$ are quite lengthy, thus we only record the results here .
It always decreasing. It produces a regular phase of
matter for $J>J_c$. Regular attendance at these non superconducting
phase has proved numerically.

\begin{figure}[h!]
\epsfxsize=7cm \epsfbox{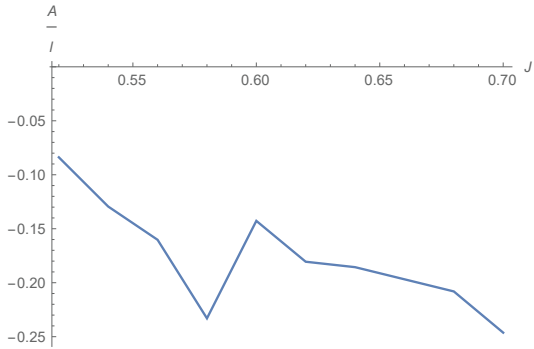} \caption{ $\frac{A}{l}$ as a function of $J$ near $\frac{1}{2}$. If system evoles in the vicinity of $J=\frac{1}{2}$, a numerical computation shows that  $\frac{A}{l}$ is a
monotonic-decreasing function.  It always decreasing. It produces a regular phase of
matter for $J>J_c$. Regular attendance at these non superconducting
phase has proved numerically.  }
\end{figure}
Formula (\ref{51}) shows a linearly-dependece expression of  reduced HEE
(\ref{HEE}) as a function total length $y_{\infty}$. The simple physical reason backs to the \emph{emergence of new extra degrees of freedom
in small values of belt length (small sizes)}. The HEE given in (\ref{HEE}) can
dominate on $y_{\infty}$ because in this limit, the main contribution
(\ref{HEE}) comes from the region $u\sim u_{*}\sim
u_{+}$. We can explain it more using the first law of thermodynamic for entanglement
entropy. It is known so far that the HEE is treated like a conventional entropy. So, it naturally
obeys the first law of thermodynamic
\cite{Bhattacharya:2012mi},\cite{Momeni:2015vka}. Let us to consider
$y_{\infty}$ as the length scale of the system. Thus, the expression
$\frac{dS}{dy_{\infty}}$ is proportional to the entangled pressure
$P_{E}=T_{E}\frac{dS}{dy_{\infty}}$  at fixed temperature in the case
of $\frac{\mu}{\mu_c}>1$. A constant slope $\frac{dS}{dy_{\infty}}$
defines a uniform entangled pressure $P_{E}$. Using  the Maxwell's
equations in thermodynamic, we know that
$\Big(\frac{dS}{dy_{\infty}}\Big)_{T}=\Big(\frac{dP}{dT}\Big)_{y_{\infty}}$.
Consequently, if we keep the
 $T$ fixed, we gain a uniform entropy gradient of
HEE $\Big(\frac{dS}{dy_{\infty}}\Big)$. This is equivalent to a
uniform  gradient of pressure $\Big(\frac{dP}{dT}\Big)$(in fixed
belt length). In this case, we see an emergent constant entropic force
\cite{Bhattacharya:2012mi}. Another simple elementary physical reason is that $S$
must be an extensive function of  characteristic size(length) of the
entangled system, namely $y_{\infty}$. From statistical mechanics we know that if we make the size of the entangled system larger, here $y_{\infty}\to k y_{\infty}$, as a vital fact, the HEE $S$ must also increases.
So, HEE $S$ must be a homogenous function of size. In our
case, $S$ is found to be homogenous  of first order, i.e.
$S(ky_{\infty})=kS(y_{\infty})$. We conclude that the $S\sim y_{\infty}$ changes
linearly with $y_{\infty}$.

  \subsubsection{ Exceeding a normal or reasonable limit: $J\rightarrow \infty$}
\begin{figure}[h!]
\epsfxsize=7cm \epsfbox{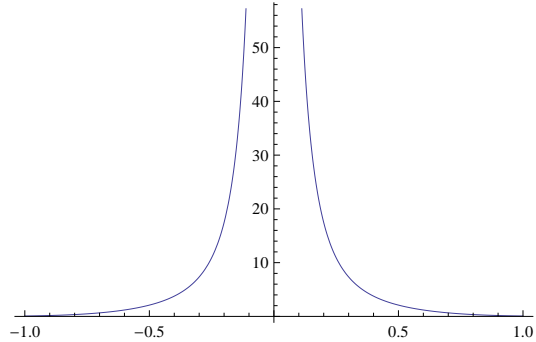} \caption{ Plot of integrand (In) of the integral (\ref{J-infinity})  as a function of $u(y_{\infty})$. If $u(y_{\infty})\to 0$, the integrand is expressed as a singular function $In:\sim \frac{0.726543}{u(y_{\infty})^2} - 0.318884 - 0.960626 u(y_{\infty})^2 $ .  }\label{integrand}
\end{figure}

The following area functional is the approximate integral form of the (\ref{func4}): When the conserved charge $J$ passes through $J=\frac{1}{2}$ at a magnitude of several orders, the functional is given by the following:

\begin{eqnarray}
&&\frac{A}{l}\asymp\frac{1}{2J}\int_{ u(-y_{\infty})}^{u(y_{\infty})}du(y)\frac{ \cosh u(y) }{\sinh^2 u(y) }\frac{\sqrt{1+2\dot{u}(y)}}{\dot{u}(y)}\label{J-infinity}
\end{eqnarray}
We plot the integrand (In) of  (\ref{J-infinity}) in figure (\ref{integrand}).

It should be noted that the solution for (\ref{ode2})  is, as a rule, only estimates, but in most instances it probably approximates closely to accuracy. This solution is given by
\begin{eqnarray}
&&\dot{u}_{\pm}(y)\equiv v+1\pm \sqrt{v^2+3v+2},\ \  v=e^{4\sinh^2u(y)}
\end{eqnarray}

\begin{figure}[htb]
\epsfxsize=7cm \epsfbox{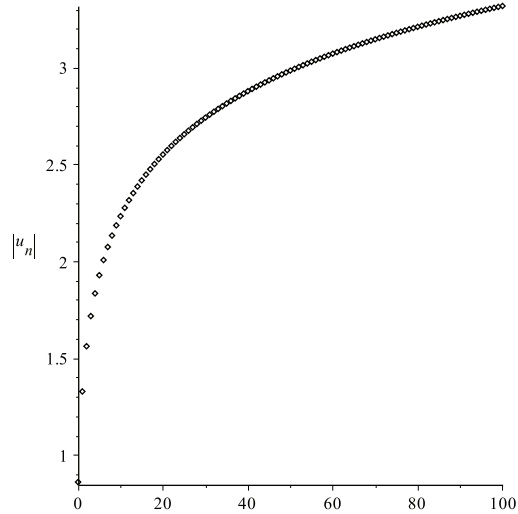} \caption{ Plot of $|u_n|$ for $n\in[0,100]$. }\label{un}
\end{figure}

This latter functional included the complex integral calculus, the calculus of improper integrals, the theory of residues and the Cauchy principal value should be found. Because $v>1$, so $\dot{u}_{-}(y)<0,\dot{u}_{+}(y)>0$. We choose the $\dot{u}_{+}(y)$ branch of the solutions to avoid a negative area functional. The method for determining exact value of the integral  (\ref{J-infinity}) is far in advance of this work, and is identical in principle with the methods of residuals. If $u(y_{\infty})\to\infty$, then the Cauchy principal value (P.V) of  (\ref{J-infinity}) is given by the following expression:

\begin{eqnarray}\label{countourintegral}
&&\frac{2JA}{l}=P.V \{\int_{ -\infty}^{\infty}In(u(y))du(y)\}=\lim_{u(y_{\infty})\to\infty}
\int_{ -u(y_{\infty})}^{u(y_{\infty})}In(u(y))du(y)\\&&\nonumber=\oint_{half} In(u(z))dz-\lim_{\epsilon\to0}\int_{ -\epsilon}^{\epsilon}In(x)du(x)-
\int_{C_{R}}
In(x)dx-\Sigma_{n=1}^{\infty}2\pi i Res\Big(In(u(z))\Big)_{u=u_n}
\end{eqnarray}

here $u_n=\sinh^{-1}\Big(\frac{\sqrt{\pi(2n+1)}}{2}e^{\frac{i\pi}{4}}\Big)   $ and $Res\Big(In(u)\Big)_{u=u_n}=\lim_{u\to u_n}(u-u_n)In(u)$ are residues of the complex function evaluated at simple(first order) poles $u=u_{n+}$. We plot the $|u_n|$ for $n\in[0,100]$ in figure (\ref{un}). We demonstrate that $|u_n|<\infty$ and never diverges. It means that the poles $u_n$ don't excess  the contour on integration with  enough large radius $R\to\infty$.

We must proof Jordan's lemma, i.e. we must prove that:

\begin{eqnarray}
|\int_{C_{R}}
In(z)dz|\leq \frac{\pi}{a}\texttt{max}_{\theta\in[0,\pi]}|g(Re^{i\theta})|,\ \ In(z)=e^{iaz}g(z).
\end{eqnarray}

In our case, $|a|=1$, $g(z)=\frac{e^{-2e^{2z}}}{\sqrt{2}}$. We have:

\begin{eqnarray}
&&\texttt{max}_{\theta\in[0,\pi]}|g(Re^{i\theta})|=\texttt{max}_{\theta\in[0,\pi]}|e^{-2\cos(2R\sin\theta)e^{2R\cos\theta}}|
\end{eqnarray}

We conclude that $ |\int_{C_{R}}
In(z)dz|_{R\to\infty}\leq |e^{-2e^{2R}}|\to 0$. So, we prove Jordan's lemma.

\begin{figure}[htb]
\epsfxsize=12cm \epsfbox{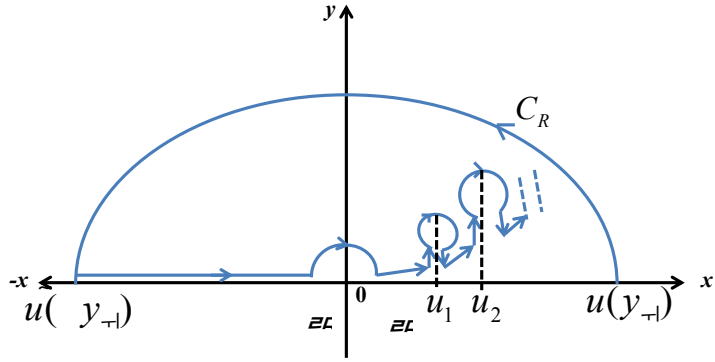} \caption{ Contour  for integration (\ref{countourintegral})}\label{contour}
\end{figure}
We compute all the necessary expressions and collect the result in the following form:

\begin{eqnarray}
&&\nonumber\frac{2JA}{l}=\Re\sum_{k=1}^{\infty}\Big[\frac{8(2k+1)^{-1}\sqrt{1+\frac{i\pi(2k+1)}{4}}}{\Pi_{n=2\neq k}^{\infty}\sinh^{-1}\Big(e^{\frac{i\pi}{4}}\Big[\frac{\sqrt{\pi(2n+1)}}{2}\sqrt{1+\frac{i\pi(2k+1)}{4}}-\frac{\sqrt{\pi(2k+1)}}{2}\sqrt{1+\frac{i\pi(2n+1)}{4}}\Big]\Big)}\Big].
\end{eqnarray}

The above expression gives us the asymptotic limit of HHE when $u(y_{\infty})\to\infty$. In this limit we see $A 	\propto\mathcal{N}\frac{l}{2J} $.

\par
As an alternative,
since $u(y_{\infty})$ is a large number, we have an approximate solution for (\ref{J-infinity}):

\begin{eqnarray}\label{A-J-inifity}
&&\frac{A}{l}\asymp\frac{1}{2J}\Big({{\rm e}^{-\frac{1}{2}\,{{\rm e}^{-2\,u(y_{\infty})}}+u(y_{\infty})}}-{{\rm e}^{-\frac{1}{2}\,{
{\rm e}^{2\,u(y_{\infty})}}-u(y_{\infty})}}-\frac{1}{2}\,\sqrt {2\pi }
{{\rm erf}\left(\frac{1}{2}\,\sqrt {2}{{\rm e}^{u(y_{\infty})}}\right)}\\&&\nonumber+\frac{1}{2}\,\sqrt {2\pi }
{{\rm erf}\left(\frac{1}{2}\,\sqrt {2}{{\rm e}^{-u(y_{\infty})}}\right)}
\Big) .
\end{eqnarray}

\begin{figure}[h!]
\epsfxsize=7cm \epsfbox{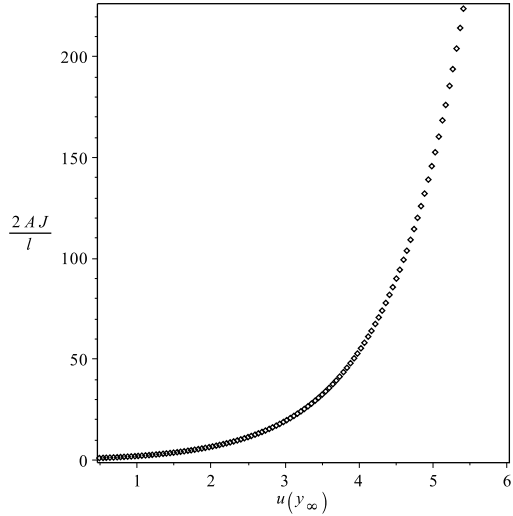} \caption{ $\frac{2 JA}{l}$ as a function of $u(y_{\infty})$ based on the estimation (\ref{A-J-inifity}). It
 shows that  HEE is a
monotonic-increasing function of $u(y_{\infty})$.  }\label{fig-A-J-inifity}
\end{figure}

We list a few results for integral in TABLE II. We know that $\log(\frac{2 JA}{l})$ is proportional to the numbers of degrees of freedom (dof) of the system $\mathcal{N}$. Just as the phenomena of sudden condensate, complete revolutions of $\mathcal{N}$ occurring to entagling appearance in a momentary interval of time, are a valid argument in favor of determinism - they may be due to the sudden emergence of new dof in system and $\mathcal{N}$ long concealed - so what looks like the orderly and necessary development of a HEE growing and exhibiting its activity in accordance with thermodynamic laws may in reality be due to the emergence of very large numbers of dof $\mathcal{N}$. There is no trace of the emergence of the new large of dofs in any other type of entropies distinct form in the HEE of time dependence backgrounds or entropic forces. We plot  the semi analytic expression (\ref{A-J-inifity}) in figure (\ref{fig-A-J-inifity}).  It
 shows that  HEE is a
monotonic-increasing function of $u(y_{\infty})$.

\begin{table}
\caption{Numerically computed values of $\log(\frac{2 JA}{l})$ in (\ref{J-infinity}) for a sample of large $u(y_{\infty})$
. Best polynomial fitting is  $ \log(\frac{2 JA}{l})=\log(64356.526- 0.152\,u(y_{\infty})+ 0.002
\,{u(y_{\infty})}^{2}
)$.}
\begin{center}
\begin{tabular}{cc}
\hline\hline
$u(y_{\infty})$& $\log( \frac{2 JA}{l})$ \\
5& $64355.86$
\\
10&$64355.17$ \\
20& $64354.47$\\
30  &$64354.07$\\
\hline\hline
\end{tabular}
\end{center}
\end{table}

\begin{figure}[h!]
\epsfxsize=7cm \epsfbox{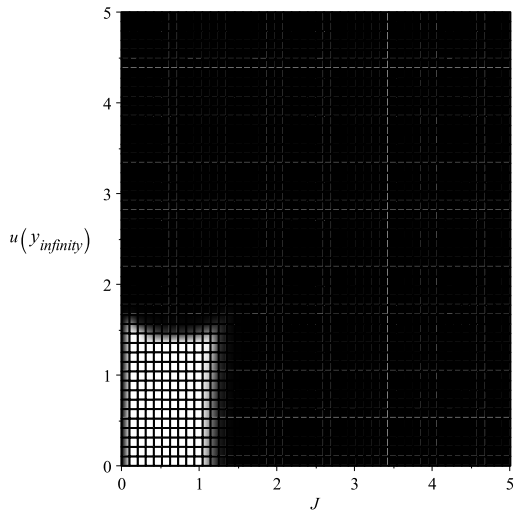} \caption{Density plot of HEE (\ref{func4}) for disconnected phase as function of $J$ and $u(y_{\infty})$.  }\label{densityplot}
\end{figure}
Furthermore, we plot density of HEE (\ref{func4}) for disconnected phase as function of $J>1/2$ and $u(y_{\infty})$ in (\ref{densityplot}). We read through the  (\ref{densityplot}) graph that, when the system  had changed from the normal phase in the disconnected phase, HEE is growing rapidly. We demonstrate that HEE is always growing with respect to time $y$. So, it doesn't violate second law of thermodynamics. It shows an additional evidence for HEE as an entropy in the sense of thermodynamics.


\par
\section{Summary}\label{Summary}
The opportunity to explore the time dependent holographic superconductors presented in several steps when we were supposing a non static AdS metric in preparation for bulk  and asked gauge fields if they wanted to go along time direction. What was the meaning of scalar condensation , with all its parade and expense, but an indirect recognition of the fact that there is effective and holographically renormalized action in the bulk to which each vacuum expectation value of dual operators is a function of time or dynamic, yet explored by authors by supposing time dependent dual chemical potential $\mu$ and charge  $\rho$densities.  But that it is easier to solve equations of motion through appropriate boundary domain and compute, in a fascinated form, with enough precision, time evolution of the dual fields. In this article we studied time dependent model for 2D holographic superconductors using $AdS_3/CFT_2$ conjecture. A time dependent version of $AdS_2$ was considered as bulk geometry. Time evolution of the gauge fields was investigated in details using numerical codes. We list the following significant results about time dependent holographic superconductors which are obtained in this work are listed as below:
\begin{itemize}
\item Gauge field and scalar field are decreasing near the AdS zone as a function of time. The decay form is exponential. System needs a long time to reach stable point.
\item For an interval of time, scalar field is diverging near the AdS zone.  We observe a discontinuity in $\psi(0,t)$ near critical point $T_c=0.01$ at time scale $t\sim 7$.  For time interval $0<t< 7, 7<t<10$,it shows that $\psi(z,t)$ is a monotonic-decreasing function.
\item We observe a discontinuity in $A_x(0,t)$ near critical point $T_c=0.01$ at time scale $t_{f}\sim 7,t_{i}\sim 0.5$. For time interval $t\in[t_i,t_f]$,it shows that $A_x(z,t)$ is a monotonic-constant function.

\end{itemize}
Furthermore we investigated holographic entanglement entropy of dual quantum systems using a generalization of the proposal of \cite{hee1}-\cite{Nishioka:2009un}. We studied extremal surfaces in two types:connected and disconnected phases. The results which we obtained are as the following:
In disconnected phase:
\begin{itemize}
\item When time coordinate $y$ is increasing, extremal function $u(y)$ in decreasing up to a local minima, after that it is increasing by a roughly slope. The minima is located at the turning point.  If we're increasing $c$, the local minima is shifted to the right. It can be interpreted as a shift in the horizon.
\item We showed that HEE is a
monotonic-increasing function.  It always increasing or remaining
constant, and never decreasing. It produces a regular phase of
matter for $T>T_c$. Regular attendance at these non superconducting
phase has proved numerically.

\end{itemize}
In disconnected phase:
\begin{itemize}
\item By a numerical computation we demonstrated that  HEE is a
monotonic-decreasing function.  It always deccreasing. It produces a regular phase of
matter for $J>J_c$. Regular attendance at these non superconducting
phase has proved numerically.

\item  We showed that $\log(S_{HEE})$ is proportional to the numbers of degrees of freedom (dof) of the system $\mathcal{N}$. Just as the phenomena of sudden condensate,  we observed an emergence of new dof in system .

\item The HEE growing and exhibiting its activity in accordance with thermodynamic laws.

\item  Density plot of HEE as function of $J$ and $u(y_{\infty})$ show that when time $y$ is growing up, for large values of $J$ the HEE reaches its maxima. So, system undergoes a normal thermodynamically time evolution according to the second law. Consequently the arrow of time never changes.
\end{itemize}


\section{Acknowledgments}
Appropriate acknowledgment of authors
 for grant provided for the program titled \emph{Regu- lar models of branes in multidimensional theories of Einstein}. H. Gholizade
thank Prof. Tapio T. Rantala for helpful discussions.

\end{document}